\documentclass[reprint,aps,amsmath,amssymb]{revtex4-2}
\usepackage{color}
\usepackage{graphicx}
\usepackage{dcolumn}
\usepackage{bm}
\usepackage[colorlinks=true,citecolor=blue,urlcolor=blue,linkcolor=blue,hyperindex]{hyperref}
\usepackage{mathtools}
\usepackage{float}
\usepackage{booktabs}
\usepackage[caption=false,justification=justified]{subfig}
\makeatother

\preprint{APS/123-QED}

\def\Tc{T_c}
\def\R2{R^2}
\def\bk{\mathbf{k}}

\makeatother
\begin{document}

\preprint{APS/123-QED}

\title{A deep learning approach to search for superconductors from electronic bands}


\author{Jun Li$^{1}$}
\email{ljcj007@ysu.edu.cn}
\author{Wenqi Fang$^{3}$}
\author{Shangjian Jin$^{2}$}
\author{Tengdong Zhang$^{1}$}
\author{Yanling Wu$^{1}$}
\author{Xiaodan Xu$^{1}$}
\author{Yong Liu$^{1}$}
\email{yongliu@ysu.edu.cn}
\author{Dao-Xin Yao$^{2}$}
\email{yaodaox@mail.sysu.edu.cn}
\affiliation{
$^1$Key Laboratory for Microstructural Material Physics of Hebei Province, School of Science, Yanshan University, Qinhuangdao 066004, China.\\
$^2$State Key Laboratory of Optoelectronic Materials and Technologies, Guangdong Provincial Key Laboratory of Magnetoelectric Physics and Devices, School of Physics, Sun Yat-Sen University, Guangzhou 510275, Peoples Republic of China. \\
$^3$Shenzhen Institute of Advanced Technology, Chinese Academy of Sciences, Shenzhen, 518055, Guangdong, China.
}

\begin{abstract}
Energy band theory is a foundational framework in condensed matter physics. In this work, we employ a deep learning method, BNAS, to find a direct correlation between electronic band structure and superconducting transition temperature ($T_c$). Our findings suggest that electronic band structures can act as primary indicators of superconductivity. To avoid overfitting, we utilize a relatively simple deep learning neural network model, which, despite its simplicity, demonstrates predictive capabilities for superconducting properties. By leveraging the attention mechanism within deep learning, we are able to identify specific regions of the electronic band structure most correlated with superconductivity. This novel approach provides new insights into the mechanisms driving superconductivity from an alternative perspective. Moreover, we predict several potential  superconductors that may serve as candidates for future experimental synthesis.
\end{abstract}
\maketitle

\section{\label{sec:intro}Introduction}

Despite the identification of over 10,000 superconductors to date \cite{m900p111d,Sommer2023}, predicting novel high-temperature superconductors remains a significant challenge. Traditionally, the discovery of superconductors has been driven by experimental synthesis efforts, followed by theoretical analyses aimed at elucidating the underlying physical mechanisms. This approach has been notably exemplified in the study of cuprate superconductors \cite{Bednorz1986} and iron-based superconductors \cite{Stewart2011}.     

In condensed matter physics, first-principles density functional theory (DFT), grounded in band theory, serves as a foundational tool \cite{Martin2004}. The electronic band structures derived from DFT calculations provide critical parameters essential for the theoretical understanding of superconductivity. These parameters are pivotal not only for conventional superconductors—such as the superconducting gap and electron-phonon interactions within BCS theory \cite{Giustino2017}—but also for unconventional superconductors, where they contribute to theories of strong correlations \cite{Aichhorn2010} and spin fluctuations \cite{Graser2010}. 

For instance, the electron-phonon coupling constant in the BCS superconductor MgB$_2$, which exhibits a relatively high superconducting transition temperatures ($T_c$) \cite{Bohnen2001}, the hydrogen-rich superconductors under high pressure that have recently set new records for $T_c$ \cite{Drozdov2019}, and two-dimension carbon-based materials \cite{Li2022,Si2013} are effectively described. 

Moreover, DFT calculations have proven instrumental in elucidating various complex phenomena in the study of unconventional superconductivity. These include nonlinear lattice dynamics in YBa$_2$Cu$_3$O$_{6.5}$ \cite{Mankowsky2014}, magnetic interactions \cite{Graser2010}, tight-binding model parameters \cite{Cao2008}, and electronic Coulomb correlations in iron-based superconductors \cite{Aichhorn2010}. Additionally, DFT has shed light on the spin-orbit coupling strength in heavy fermion systems \cite{Samokhin2004}, interlayer interactions in bilayer twisted graphene \cite{Carr2018}, $\sigma$-bond metallization under high pressure in nickel-based superconductors \cite{Luo2023}, as well as the superconducting pairing symmetry in bilayer silicene \cite{Liu_2013} and two-dimensional carbon-based materials \cite{Li2020,Ye2023}.

\begin{figure*}[t]
\centering
\includegraphics[width=0.75\textwidth]{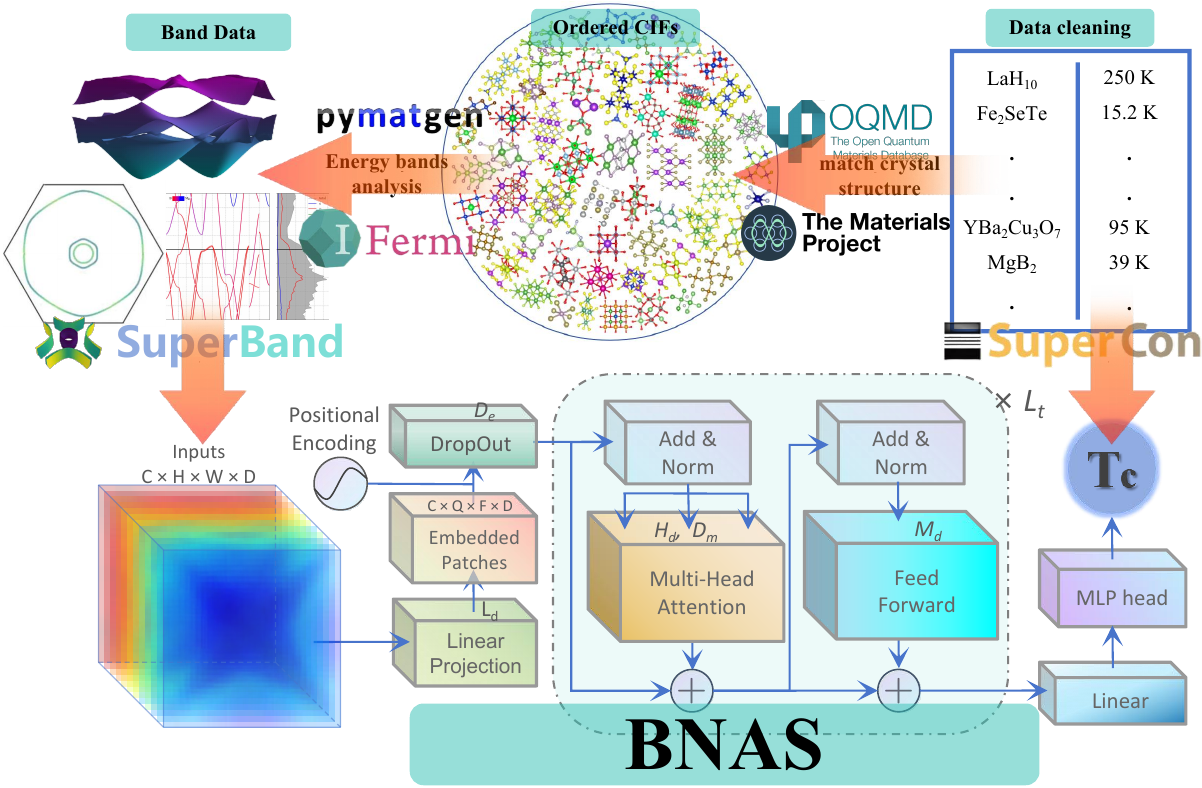}
\makeatletter\long\def\@ifdim#1#2#3{#2}\makeatother
\caption{\label{fig:2}Schematics of the deep learning method. Chemical formulas and $T_c$ data are sourced from the SuperCon database \cite{m900p111d}, which contains duplicate entries and is addressed by a data cleaning process. CIF files are retrieved from the Materials Project (MP) \cite{MPS} and the Open Quantum Materials Database (OQMD) \cite{Kirklin2015}, but many contained disordered configurations. An order transformation method from 3DSC methodology \cite{Sommer2023} is used to retain ordered structures with the lowest Ewald energy. High-throughput calculations are conducted via the Atomate package \cite{Mathew2017}, with parameters from the MIT High-Throughput Project \cite{Jain2011}, and workflow automation managed by FireWorks \cite{CPE:CPE3505}. The Pymatgen package \cite{Jain2013,Ong2015} and Ifermi package \cite{Ganose2021} are used to extract band structure data with dimensions C $\times$ L $\times$ H $\times$ W = 18 $\times$ 32 $\times$ 32 $\times$ 32 to form the Superband Dataset. In the BNAS software, we implement a 3D Vision Transformer (3D-ViT) \cite{dosovitskiy2020vit} to process band data, using a set of optimal hyperparameters P $\times$ Q $\times$ F $\times$ D = 18 $\times$ 8 $\times$ 8 $\times$ 8, $L_d$ = 534, $D_e$ = 0.127, $H_d$ = 64 , $D_m$ = 0.197, $M_d$ = 1038, and $L_t$ = 3. }
\end{figure*}

It will be useful to use the machine learning (ML) to find the underlying relations between the electronic bands by DFT and the experimental $T_c$ values.
Recently, ML has garnered significant attention in the study of superconductivity, leading to several innovative applications. These include binary classification based on a $T_c$ threshold of 10 K \cite{Stanev2018}, deep learning models designed to interpret chemical formulas within the framework of the periodic table \cite{Konno2021}, and hybrid neural networks that combine convolutional neural networks (CNNs) with long short-term memory networks (LSTMs) for advanced analysis \cite{Li2020d}. Additionally, closed-loop ML methods have been employed for the rapid exploration of vast material search spaces \cite{Pogue2023,Babic2023}, while studies on iron-based superconductors have utilized lattice parameters as key inputs \cite{Hu2021}. Other noteworthy contributions include research on cuprate superconductors \cite{Wang2023a}, characterization of Eliashberg spectral functions \cite{Xie2022a}, and improvements in Allen-Dynes fitting \cite{Xie2019}. Analyses based on atomic vectors and symmetries, as well as atomic position smooth overlap and crystal structure \cite{Zhang2022}, have also been explored, alongside investigations into the correlation between Debye temperature and $T_c$ \cite{Smith2023}.

Most prior ML studies on superconductivity have primarily focused on the chemical composition or specific physical properties of superconductors. While these approaches have demonstrated strong predictive performance for known superconductors, they often lack theoretical groundings necessary for the discovery of novel superconductors. It is important to note that isomers with identical chemical compositions can exhibit significantly different electronic band properties, and materials with distinct elemental compositions may share similar electronic band structures, leading to analogous electrical properties. Thus, electronic band data provide a more fundamental and intuitive understanding of superconducting phenomena.

In this paper, we construct a training dataset consisting of 1,362 experimentally validated superconductors and 1,112 non-superconducting materials for ML applications. We then develop a deep learning framework named BNAS capable of predicting the $T_c$ of superconductors without relying on empirical parameters. The incorporation of an attention mechanism within the deep learning model allows for the identification of specific electronic band regions most strongly associated with superconductivity. Furthermore, analysis based on these superconductors's electronic bands has yielded a wealth of statistically significant insights, which are invaluable for elucidating the underlying mechanisms of superconductivity, particularly in the context of high-temperature superconductors.

\section{Methods}\label{sec:method}
\begin{figure*}[t]
\centering
\includegraphics[width=0.5\textwidth]{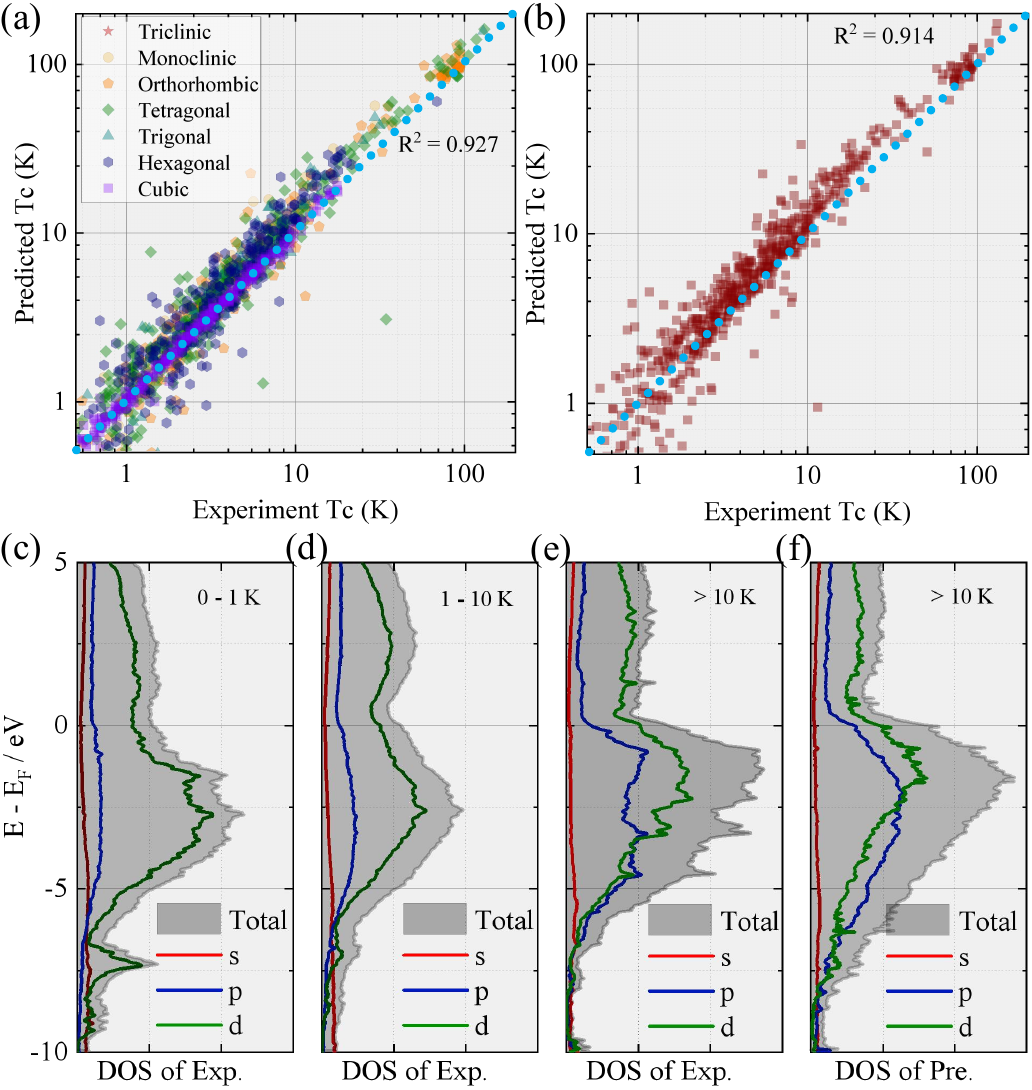}
\captionsetup[figure]{justification=justified, singlelinecheck=off} 
\caption{\label{fig:1}
The accuracy of BNAS in predicting superconducting properties is demonstrated as follows: (a) shows comparison between the BNAS-predicted $T_c$ and experimentally measured $T_c$ for the training set after removing duplicates from data augmentation, with $R^2$ of about 0.927. Non-superconductors, for which $T_c$ is set to 0, contribute minimally to the $R^2$ calculation, and thus are not included in the analysis of the training set. (b) shows comparison of BNAS-predicted $T_c$ with experimentally measured $T_c$ for the validation set, which was not involved in neural network training. An $R^2$ of about 0.918 is observed, again excluding non-superconductors. (c) shows the average DOS for 206 experimental superconductors with $T_c < 1$ K, while (d) presents the average DOS for 967 experimental superconductors with $T_c$ in the range of 1-10 K. (e) illustrates the average DOS for 189 experimental superconductors with $T_c > 10$ K. (f) shows the average DOS for 1,502 superconductors identified by BNAS as likely to have $T_c$ greater than 10 K, based on large-scale searches.}
\end{figure*}

\begin{figure*}[t]
	\centering
	\includegraphics[width=0.7\textwidth]{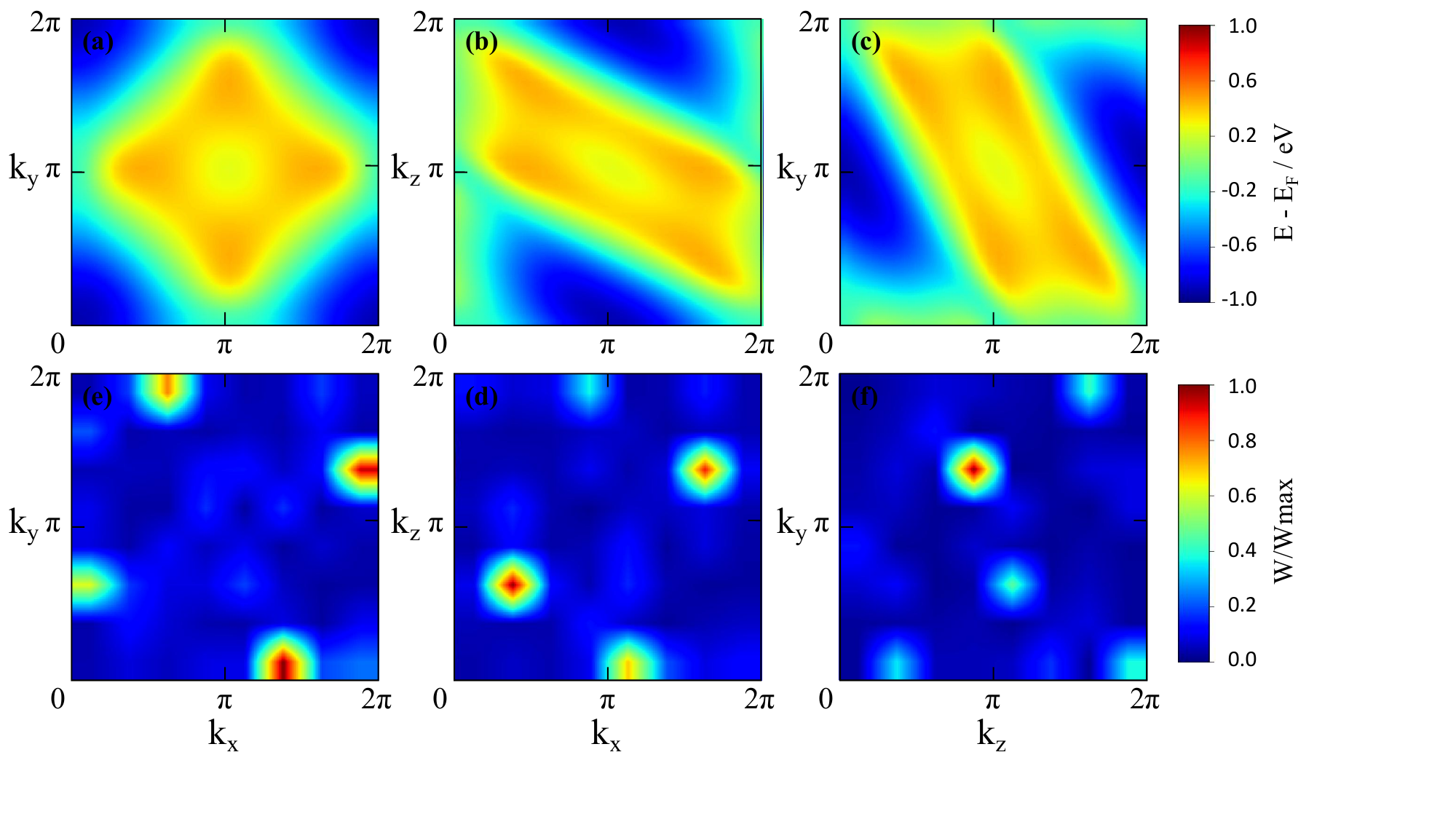}
	\caption{\label{fig:3}The attention mechanism in the deep learning model for KFe$_2$Se$_2$: the electronic band structure around the Fermi surface (a) k$_x$-k$_y$ plane, (b) k$_x$-k$_z$ plane, and(c) k$_z$-k$_y$ plane. The attention mask weights (d) k$_x$-k$_y$ plane, (e) k$_x$-k$_z$ plane, (f) k$_z$-k$_y$ plane, as determined using the Attention Rollout method \cite{abnar-2020}. The selected electronic band crosses the Fermi level. In the attention mask weight visualization, red areas indicate regions with a greater contribution to superconductivity, whereas blue areas represent regions with negligible effect on superconductivity.}
\end{figure*}

\textit{Dataset generation.} To construct electronic band structures for ML training set, we extract the $T_c$s and DFT results of 1362 superconductors and 1112 non-superconducting compounds from Superband Dataset \href{http://www.superband.work/}{http://www.superband.work/}. The reported $T_c$ and 18 electronic bands near the Fermi level are chosen for the training. Each electronic band is coded in a $32\times 32\times 32$ $\bk$ grid with the energies defined related to the Fermi level, resulting in band data of one material with dimensions C (channel) $\times$ L (length) $\times$ H (height) $\times$ W (width)= 18 $\times$ 32 $\times$ 32 $\times$ 32. Since the number of superconductors decreases exponentially with the transition temperatures $T_c$, it is not an ideal metric for training. Therefore, we use $\log(T_c/K + 1)$ ($K$ corresponding to Kelvin) as the training metric.


The training set of 2,474 is insufficient for deep learning training. Therefore, data augmentation techniques must be employed to expand the training set in order to avoid overfitting. Firstly, since the k-space coordinates (k$_x$, k$_y$, k$_z$) of the electronic band data are defined relative to the z-axis and are therefore interchangeable, these three dimensions can be permuted. Secondly, the real-space atomic coordinates of the material are swapped (e.g. k$_x$ $\rightarrow$ -k$_x$) without changing the intrinsic properties of the material. This transformation retains all relevant information in the electronic band data. Using these data augmentation strategies, the training set is expanded to 59,376 data points.  

\textit{Deep learning algorithms.} We develop our deep learning model to the BNAS (Fig.~\ref{fig:2}). Due to the similarities of the electronic band, we utilize the three-dimensional Vision Transformer (3D-ViT) \cite{dosovitskiy2020vit} from the Transformer deep learning architecture. The input data for the training thus contains dimensions C $\times$ L $\times$ H $\times$ W = 18 $\times$ 32 $\times$ 32 $\times$ 32. Before the deep learning process, a linear projection layer is employed as the input layer to uniformly process the band data, with the dimensionality controlled by a positive integer $L_d$. Following this, a patch embedding layer is used to rearrange the processed band data, with dimensions specified as C $\times$ Q (patch length) $\times$ F (patch height) $\times$ D (patch width) = 18 $\times$ 8 $\times$ 8 $\times$ 8. To mitigate overfitting, a dropout layer is incorporated to randomly select and discard a portion of the data, with the dropout rate, $D_e$, ranging between 0 and 1.

We now summarize our deep learning in the light blue box in Fig.~\ref{fig:2}. A transformer network with $L_t$ layers is constructed. Each transformer layer includes a multi-head attention mechanism, with the number of attention heads controlled by $H_d$ and the dropout rate of the attention output controlled by $D_m$. A feedforward neural network, with its dimension controlled by $M_d$, provides additional processing for the attention mechanism. Here, $H_d$, $D_m$, and $M_d$ are adjustable positive integers. The network concludes with a linear layer and a multilayer perceptron (MLP) head as the output layers, producing a one-dimensional tensor as the final output, which is the predicted $T_c$.

To train the model, we use the Optuna~\cite{optuna_2019} to automate the search for optimal hyperparameters, employing the mean squared error (MSE) between the predicted outputs and the $\log(T_c/K + 1)$ values of the training set as the loss function. A large-scale parameter search is conducted over 100 epochs to identify the deep learning parameters that minimize the loss function. The optimal hyperparameters determined are $L_d$ = 534, $D_e$ = 0.127, $H_d$ = 64 , $D_m$ = 0.197, $M_d$ = 1038, and $L_t$ = 3, with further details provided in Appendix \ref{sec:tra}. The dataset is split into training and validation sets in a 4:1 ratio for cross-validation. Ultimately, we obtain a well-trained deep leaning model.

\section{\label{sec:results}Results and discussion}
\begin{figure*}
	\centering
	\includegraphics[width=0.75\textwidth]{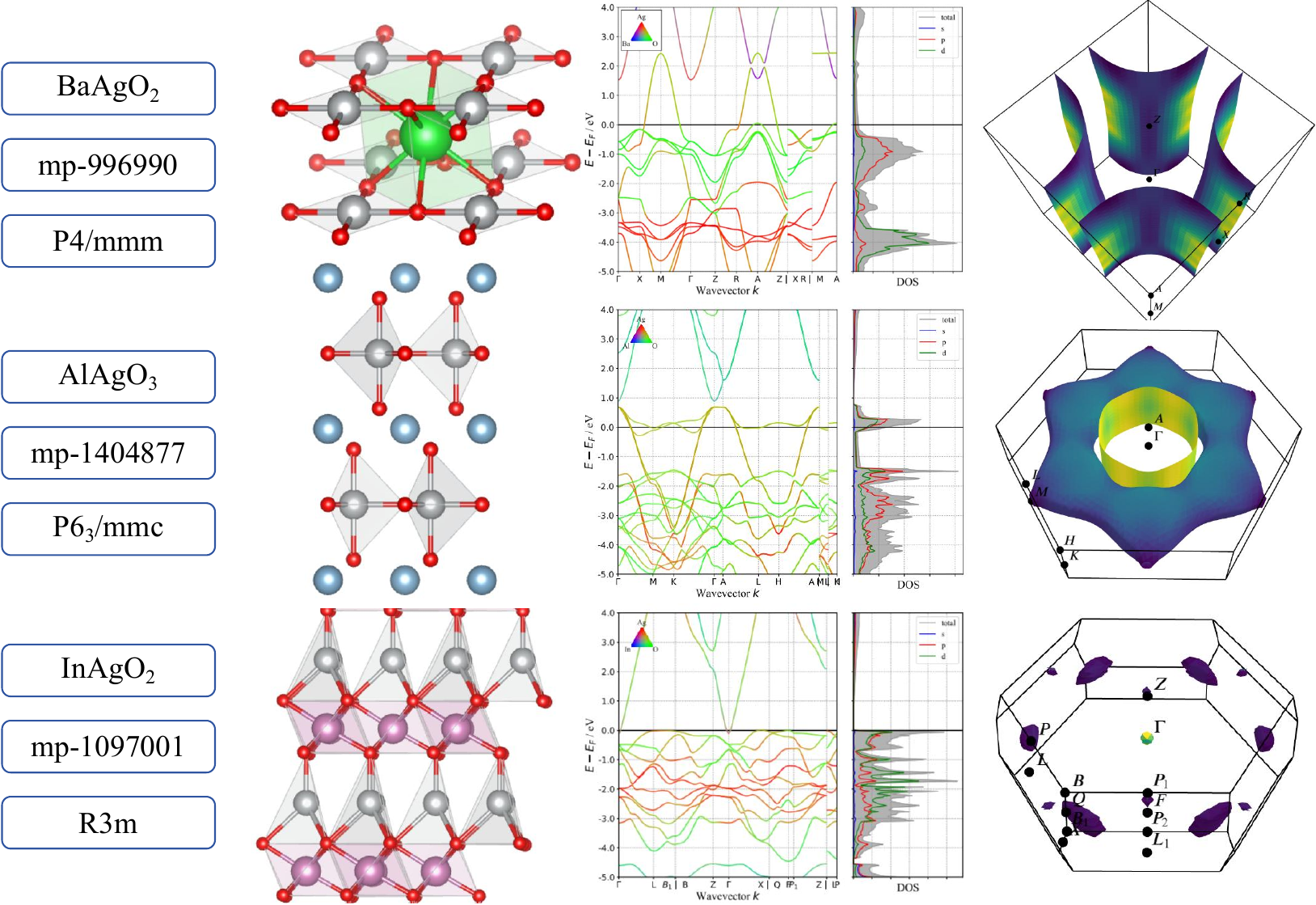}
	\caption{\label{fig:4}The three candidates for noble metal high-temperature superconductor, BaAgO$_2$ (mp-996990), AlAgO$_3$ (mp-1404877), and InAgO$_2$ (mp-1097001). For each material, we provide space group, crystal structure, electronic band structure, and Fermi surface, as sourced from the MP \cite{MPS}. }
\end{figure*}

To show the validity of our deep learning model, we compare the predicted $T_c$ with the experimental values for the known superconductors (Fig. \ref{fig:1}). The goodness of predicted $T_c$ is quantified using the coefficient of determination, $R^2$, defined by:
\begin{equation}\label{eq-r2}
R^2=1-\frac{S_{\text {Res}}}{S_{\text {Tot}}}=1-\frac{\sum_i\left(T_i-\hat{T}_i\right)^2}{\sum_i\left(T_i-\bar{T}\right)^2}, 
\end{equation}
where $T_i$ represents the predicted $T_c$, $\hat{T}_i$ denotes the average of predicted $T_c$, and $\bar{T}$ denotes  averaged $T_c$. The drawback of MSE lies in its nature as an absolute error metric. While it can be effectively used as a loss function for neural network training, the values are not easily interpretable, especially in cases where we use $\log(T_c/K + 1)$ as the label. The $R^2$ metric offers a standardized, intuitive, and relative measure of model performance.


For training set, we get the $R^2=0.927$. For validation set (not used in training), the model's predictions provide good agreement with the experimental superconductors, giving an $R^2=0.918$. The accuracy is particularly high for superconductors with cubic crystal structures. We observe that some $T_c$ predictions with hexagonal and tetragonal systems are approximately 20$\%$ higher than the experimental $T_c$, which is not attributable to inadequate training steps. Our deep learning model has reached the lowest saturation of the loss value, as detailed in Appendix \ref{sec:tra}. This discrepancy suggests that the real $T_c$s of these materials may be higher than currently observed experimentally. Overall, the deep learning approach demonstrates high efficiency in correlating electronic bands with $T_c$, indicating that electronic bands alone can serve as effective indicator of superconductivity.

\begin{table}
    \caption{\label{tab:1}Benchmark of predicted $\Tc$ vs. experimental $\Tc$ of four pairs of non-superconducting materials and superconductors.  }
    \begin{tabular}{l|lcc}
    \hline \hline
      Structure & System & Exp. $\Tc$ & Pre. $\Tc$  \\ \hline
        Ba$_2$Cu$_4$PrO$_8$ \cite{Yamada1994}  
        & Cmmm &  - & $<$ 0.01 K\\
        Ba$_2$Cu$_4$TmO$_8$ \cite{Hijar1995} 
        & Ammm &  80 K & 103.4 K\\
        LaYAs$_2$Fe$_2$O$_2$ \cite{Tropeano2009}  
        & P4mm & -  & $<$ 0.01 K\\
        LaAsFeO \cite{Okada2008} 
        & P4/nmm & 10.2 K  & 12.6 K \\ 
        La$_3$Ni$_2$O$_7$ \cite{Sun2023}  
        & Amam & - & 0.45 K\\
        La$_3$Ni$_2$O$_7$ 29.5 GPa\cite{Sun2023}
        & Fmmm & 80 K  & 116.3 K \\ 
        IrSiTb \cite{Szytula2001}
        & Pnma & - &  $<$ 0.01 K\\ 
        IrSiZr \cite{Suzuki2016}  
        & Pnma & 2.04 K  & 1.79 K\\
        \hline \hline
    \end{tabular}
\end{table}
We illustrate the high efficiency of our deep learning model using four examples of superconductors, as detailed in Tab. \ref{tab:1}. The impact of doping on cuprate superconductors is significant, for instance, Ba$_2$Cu$_4$PrO$_8$ \cite{Yamada1994} and Ba$_2$Cu$_4$TmO$_8$ \cite{Hijar1995} have similar electronic band structures, yet the former is not a superconductor while the latter exhibits $T_c = 80$ K. 
This discrepancy is accurately captured by our deep learning model.

In iron-based superconductors, the role of magnetism is crucial, and the challenge of predicting superconductivity is compounded by the distinct electronic bands for spin-up and spin-down states. Our model effectively find the difference between the superconductivity of LaYAs$_2$Fe$_2$O$_2$ \cite{Tropeano2009} and the non-superconducting LaAsFeO \cite{Okada2008}, both of which are magnetic.

For nickel-based superconductors, which exhibit superconductivity under pressure, our model successfully verifies the found superconductivity in La$_3$Ni$_2$O$_7$ under pressure \cite{Sun2023} and confirms its non-superconductivity without pressure.

Additionally, the electronic bands of alloys are highly complicated. Our model distinguished superconducting IrSiTb \cite{Szytula2001} and non-superconducting IrSiZr \cite{Suzuki2016}, even though both share a similar electronic band structure within the same Pnma space group. These examples demonstrate the robustness and general applicability of our deep learning model in determining superconductivity across various materials.

Using the Attention Rollout method \cite{abnar-2020} within the 3D-ViT framework, we identify the regions of the electronic band structure most closely associated to superconductivity. We illustrate this with the example of KFe$_2$Se$_2$ \cite{Ying2012}. Fig. \ref{fig:3} shows the electronic band structure near the Fermi surface of KFe$_2$Se$_2$, along with the corresponding attention mask. KFe$_2$Se$_2$ crystallizes in the space group I4/mmm, which is reflected in the C$_4$ symmetry observed in the k$_x$-k$_y$ plane of the electronic band structure and a tilt C$_2$ symmetry in the k$_z$-direction. The attention mask for the x-y plane also clearly displays C$_4$ symmetry, with regions identified by the deep learning model as contributing significantly to superconductivity appearing near (0, $\frac{4\pi}{3}$). In the z-direction, two regions of notable weight are observed near $k_z = \frac{4\pi}{5}$ and $\frac{6\pi}{5}$. Despite the initial randomization of model weights without considering material symmetry, the trained deep learning model effectively captures symmetry-related features in the attention mask. This highlights the capability of deep learning to analyze 3D electronic bands. The detail of the  attention method and additional examples are provided in Appendix \ref{sec:att}. 

Using our deep learning model, we conduct a comprehensive search within the MP\cite{MPS} database for materials with energy gaps smaller than 0.2 eV. This screening process involves evaluating 46,442 materials. This search finds 14,956 candidates with potential $T_c$ above 2 K, among which 1,502 candidates could have $T_c$ exceeding 10 K. Details are presented in Appendix \ref{sec:pre}. 

Next, we analyze the density of states (DOS) of superconductors. As shown in Fig. \ref{fig:1}(c)(d)(e), we present the average DOS for all experimentally confirmed superconductors. Our analysis reveals that s-orbitals contribute minimally to superconductivity, while p-orbitals and d-orbitals play more significant role. Notably, materials with $T_c > 10$ K generally contain transition metal elements, indicating that a higher $T_c$ is associated with greater contributions from d-orbitals.

We observe a pattern where the DOS near the Fermi surface is relatively low, with high DOS values both below and above the Fermi level, and a sharp decrease near the Fermi surface. This trend suggests that a higher $T_c$ corresponds to a steeper slope in the DOS near the Fermi level. This pattern is also reflected in the predicted DOS averaged for the 1,502 materials likely to have $T_c$ greater than 10 K, as shown in Fig. \ref{fig:1}(f). These findings further validate the efficiency of our deep learning model and show the potential for discovering new superconductors.

Among the identified materials, we highlight three promising  superconductors with noble metal: BaAgO$_2$ (mp-996990), AlAgO$_3$ (mp-1404877), and InAgO$_2$ (mp-1097001). These materials belong to different space groups. BaAgO$_2$ exhibits a DOS minimum near the Fermi surface, with a maximum $\sim$ 0.2 eV below the Fermi level. Its band structure reveals a pronounced peak at the ($\pi$, $\pi$) point, resulting in a nearly circular Fermi surface, reminiscent of the characteristics of cuprate superconductors. Although BaAgO$_2$ has yet to be experimentally synthesized, theoretical studies suggest its kinetic and thermodynamic stability \cite{Cerqueira2015}, making it a candidate high-temperature superconductors.

AlAgO$_3$ features an almost flat band near the Fermi surface ($\pm$  0.2 eV) and a very high DOS at the Fermi level. Its Fermi surface also shows nesting, which is favorable for unconventional superconductivity. While InAgO$_2$ does not exhibit significant band overlap at the Fermi surface, it has a high DOS value just below the Fermi level ($<$ 0.1 eV), indicating potential superconductivity.

Notably, replacing Ag with Au in these materials results in similar electronic band structure. For instance, in BaAuO$_2$ (mp-1147676), there is potential for discovering new type of unconventional superconductors. More potential superconductors are shown in Appendix \ref{sec:pre}. 

\section{\label{sec:con}conclusion}
We employ deep learning methods to try to establish a correlation between electronic bands and T$_c$. Our findings indicate that electronic bands can serve as indicators of superconductivity. To mitigate overfitting, we utilize a relatively simple deep learning neural network model. Despite its simplicity, this model effectively tests and predicts superconducting properties. Using the attention mechanism inherent in deep learning, we successfully identify specific regions of the electronic band structure most closely associated with superconductivity. This approach provides a new way for understanding the underlying mechanisms of superconductivity. Additionally, our method enables the finding of new superconductors, which can be experimentally realized in future.


This study presents a novel approach to electronic band analysis and proposes a new framework for predicting the electrical properties of materials. As the developing of artificial intelligence and high-throughput DFT technologies, the difficulties related to neural network model limitations and training dataset bottlenecks discussed in this paper are likely to be overcome. The integration of ML with electronic band analysis, as introduced here, will be a possible new direction for AI in materials.


\appendix

\section{\label{sec:tra}Model training}
Studies on both conventional and unconventional superconductors indicate that the electronic band density near the Fermi surface greatly influences superconductivity, whereas bands far from the Fermi surface contribute little \cite{Li2020,Ye2023}. 
However, the number of electronic bands near the Fermi surface varies across different superconductors. While ML can indeed take the superconductor with the highest number of electronic bands as a reference and pad the bands of superconductors with fewer electronic bands with zero tensors to form a training set, this approach leads to a waste of computational resources and introduces additional uncertainties in deep learning.

\begin{table}
        \caption{\label{tab:2}The deep learning performance of hyperparameters on the MSE validation loss after 100 epochs of training. }
    \begin{ruledtabular}
    \begin{tabular}{c|cccccc}
    Validation loss&$L_t$& $L_d$&$H_d$&$M_d$&$D_e$&$M_d$\\ \hline
    0.324&7&734&24&2154&0.457&0.458\\
    0.240&4&82 &32&246 &0.283&0.360\\
    0.408&8&618&8 &2346&0.204&0.266\\
    0.346&8&468&16&2290&0.412&0.223\\
    0.313&6&962&64&2796&0.461&0.438\\
    0.248&7&120&36&3810&0.415&0.294\\
    0.239&5&754&16&1372&0.268&0.451\\
    0.308&7&188&22&3298&0.161&0.189\\
    0.363&3&114&46&226 &0.172&0.241\\
    0.266&4&912&24&880 &0.149&0.327\\
    0.313&6&562&48&482 &0.273&0.141\\
    0.340&7&558&64&306 &0.492&0.240\\
    0.277&4&334&14&3478&0.289&0.157\\
    0.228&3&534&64&1038&0.127&0.197\\
    \end{tabular}
    \end{ruledtabular}
\end{table}

\begin{figure}
\centering
\includegraphics[width=0.45\textwidth]{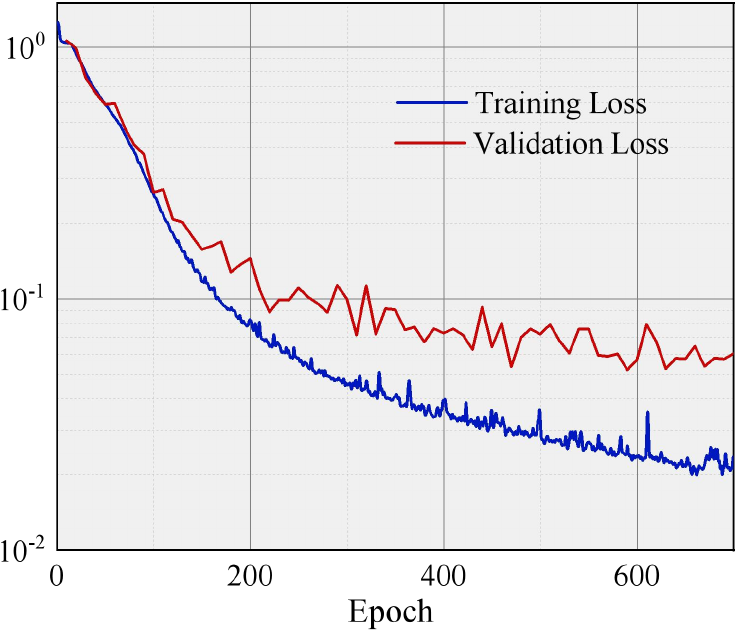}
\caption{\label{fig:a1}During the actual training process, with the given hyperparameters, the model iterates over all data points in the training set in a random order for each epoch, calculating the training loss. Every 10 epochs, the validation loss is computed using all data points from the validation set. After 400 epochs, the validation loss reaches saturation, even though the training loss continues to decrease. Further training beyond this point would offer no meaningful improvement, so the final model is selected after 700 epochs.}
\end{figure}

\begin{figure*}[t]
\centering
\includegraphics[width=0.75\textwidth]{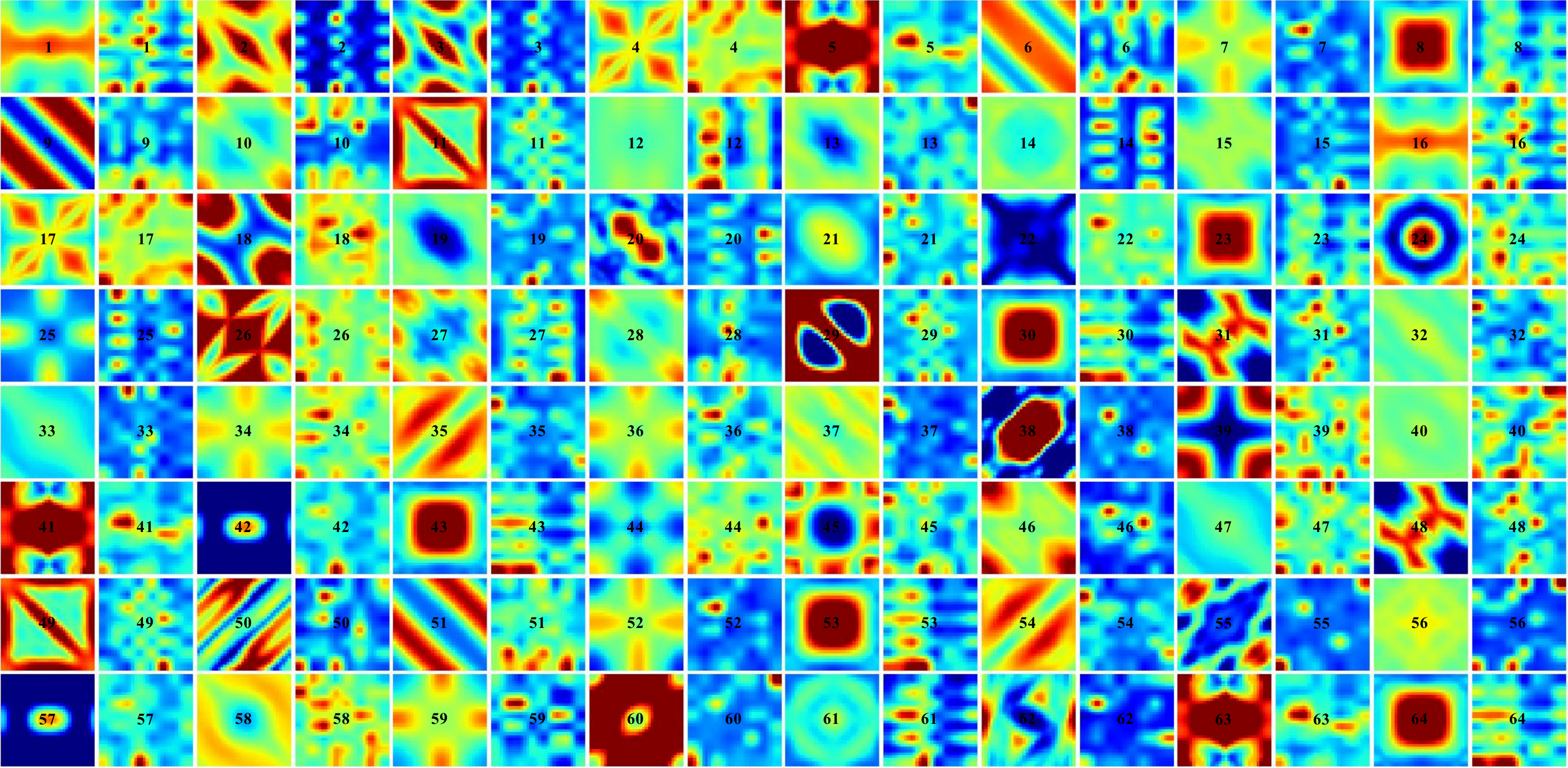}
\caption{\label{fig:a3}Further example attention weights as in Fig. \ref{fig:3}(a)(d), we have only selected the k$_x$-k$_y$ plane, as the z-axis is typically set as the principal axis, and the k$_x$-k$_y$ plane tends to exhibit the highest symmetry. The examples shown here are randomly selected from those with $T_c$ above 5 K.}
\end{figure*}

\begin{figure*}
\centering
\includegraphics[width=0.95\textwidth]{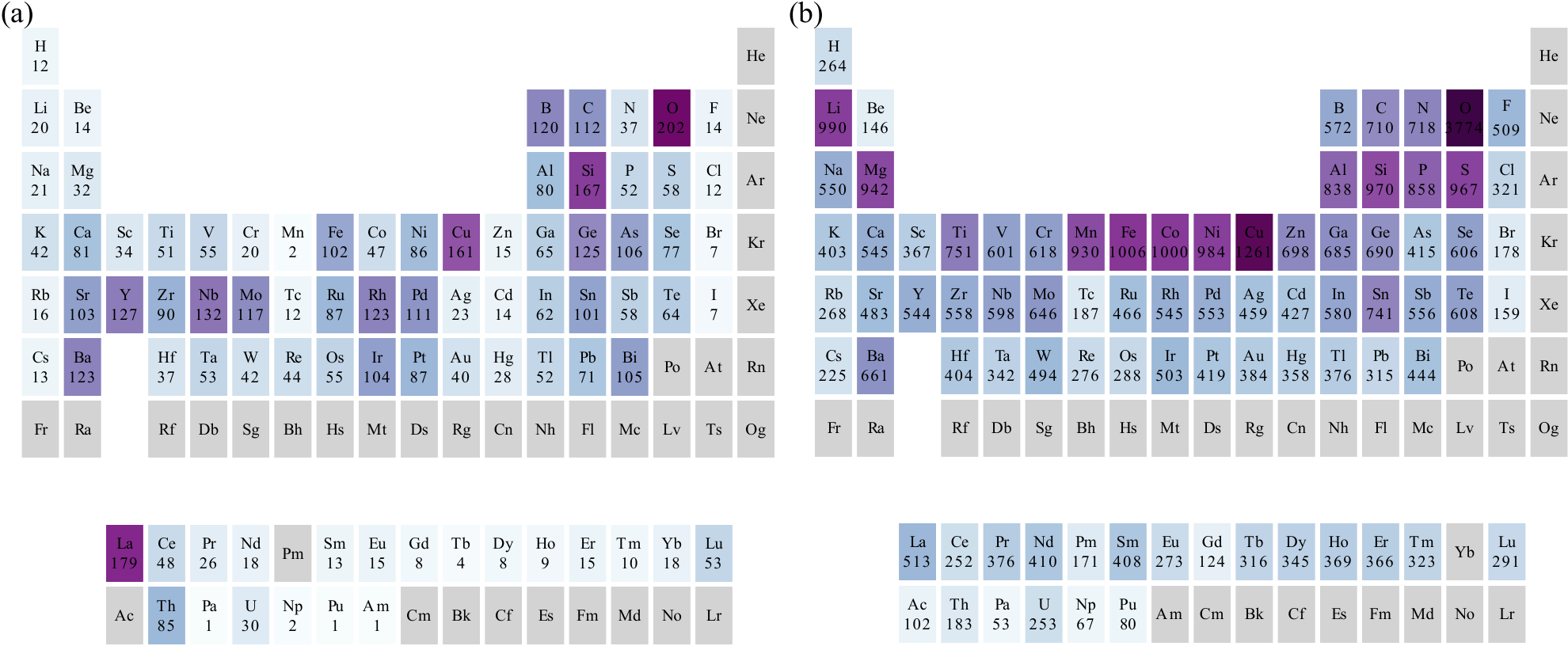}
\caption{\label{fig:a2}The elemental distribution of (a) superconductors found experimentally and (b) predicted superconductors. For any superconductor containing element A, the count for element A is incremented by 1. By following this process, we obtain the elemental distribution across all superconductors. It is important to note that this analysis encompasses all superconductors, without distinguishing between conventional and unconventional ones. The abundance of metallic elements primarily originates from early alloy-based superconductors. Additionally, oxygen is present not only in cuprate superconductors but also in many low-temperature superconductors.}
\end{figure*}
Based on our analysis, the average energy of the 18 bands around to the Fermi surface for most superconductors is typically within the range of -3 eV to 3 eV. Consequently, we uniformly select 18 bands near the Fermi surface for our study. Concerning the grid resolution of the band data, a denser spatial grid can certainly yield more accurate results but at the expense of increased storage and computational resources. In DFT calculations, the typical reciprocal space grid resolution for band structures is generally below 12 $\times$ 12 $\times$ 12. Expanding to a higher grid resolution during data augmentation could potentially introduce data distortion issues. Therefore, after careful consideration, we select an 18 $\times$ 32 $\times$ 32 $\times$ 32 grid as the dimensionality for the band data, aiming to find a balance between accuracy and computational efficiency.

It is well known that Vision Transformers (ViTs) typically utilize a patch size of 16 $\times$ 16 when processing two-dimensional images \cite{dosovitskiy2020vit}. For our 3D data, we adapt a resolution 8 $\times$ 8 $\times$ 8 of for each image patch, resulting in a total of 4 $\times$ 4 $\times$ 4 patches. Additionally, we employ the Optuna software \cite{optuna_2019} to optimize the hyperparameters of our model, using validation loss as the optimization criterion. The performance of the selected hyperparameters, following 100 epochs of training, is summarized in Table \ref{tab:2}.

Interestingly, our observations indicate that increasing the number of deep learning parameters does not necessarily improve the results. For example, a lower transformer layer count ($L_t$) yields superior training outcomes compared to models with higher $L_t$ values. We suppose that an excessive number of parameters might lead to overfitting, which explains why the hyperparameters set with the lowest validation loss, as presented in Table \ref{tab:2}, is selected.

To train the model, we utilize stochastic gradient descent (SGD) with a learning rate of 0.001, momentum of 0.9, weight decay of $10^{-5}$, and batch size of 64. The training process is depicted in Fig. \ref{fig:a1}, where we observe that although the training loss continued to decrease slightly up to 400 epochs, the validation loss plateaued at around 0.06. Therefore, we extend the training to 700 epochs to finalize the deep learning model.

\section{\label{sec:att}Attention mechanism}

\begin{figure*}
\centering
\includegraphics[width=0.75\textwidth]{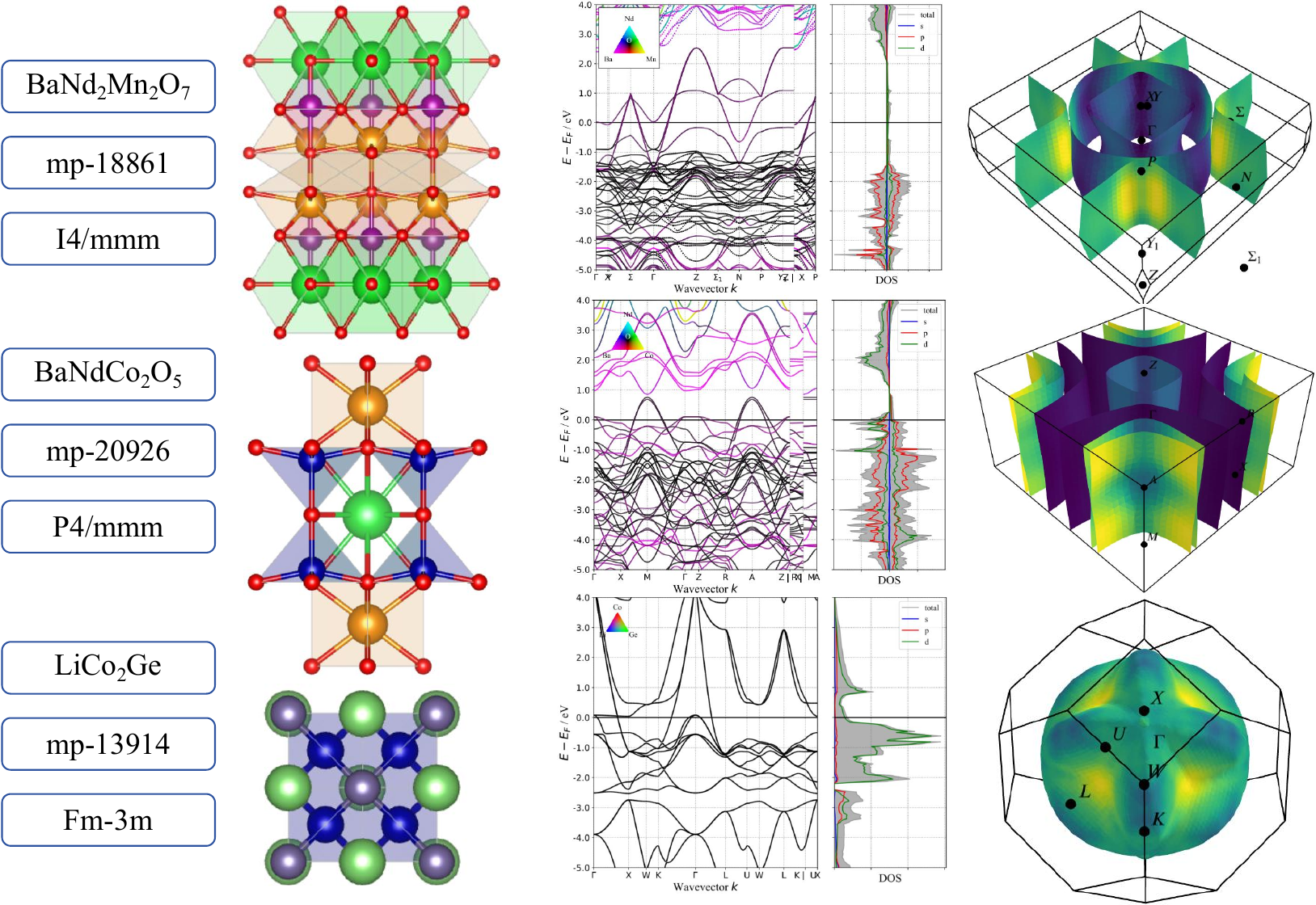}
\caption{\label{fig:a4}Further example as in the left of Fig. \ref{fig:4} with three possible manganese, cobalt and lithium-based superconductor candidates.}
\end{figure*}

To generate attention maps from the output token to the input space, we employ the Attention Rollout method \cite{abnar-2020}. In summary, we average the attention weights across all heads in the 3D-ViT model and recursively multiply the weight matrices across all layers. This method effectively captures the cumulative effect of attention across tokens throughout the entire network, providing a comprehensive view of how information is propagated and emphasized from input to output.

The Attention Rollout method offers an intuitive approach to trace the flow of information as it propagates from the input layer through the higher layers of a neural network. By systematically aggregating and multiplying attention weights across layers, this technique enables us to visualize how different parts of the input influence the model's decision-making process at various stages. Attention Rollout provides a detailed representation on how the model integrates and refines information as it progresses through each layer, ultimately leading to the final output. This makes it an invaluable tool for dissecting the inner workings of 3D-ViT models, offering insights into how these models interpret and prioritize various aspects of the input data.

Further examples of attention weights are shown in Fig. \ref{fig:a3} (random selection within the k$_x$ - k$_y$ plane). While some images may appear blurred, the majority of attention weights successfully captures the crystalline symmetry of the materials. We believe that as this technology advances, these attention weights will help identify the key features of electronic bands that contribute to superconductivity, providing a reliable reference for addressing the challenges of high-temperature superconductivity mechanisms. Additionally, our model can also use theoretical lattice models, such as tight-binding models, as inputs, aiding in determining whether these lattice models contain critical factors conducive to the formation of superconductivity.

\section{\label{sec:pre}Superconductor Candidates}

Early superconductors were primarily derived from metals or alloys, which inherently possess conductive properties. In contrast, unconventional superconductors, apart from those that naturally exhibit conductivity, often emerge from strongly correlated systems like Mott insulators, with no band gap. Therefore, when identifying potential superconductor candidates, our focus is primarily on materials with no band gap or possess a slightly doped gap. To this end, we systematically screened 46,442 materials in MP database \cite{MPS} with band gaps up to 0.2 eV to find possible superconductor candidates.

We employ the trained deep learning model to conduct high-throughput computational analysis of the band structures of these materials, ultimately identifying 14,956 potential superconductors. The elemental distribution of these candidates is depicted in Fig. \ref{fig:a2}(b), and is compared to the distribution of experimentally confirmed superconductors in Fig. \ref{fig:a2}(a).

We observe that among the predicted superconductors, the fourth-period transition metals—manganese (Mn), iron (Fe), cobalt (Co), nickel (Ni), and copper (Cu)—appear with notable frequency, all of which are adjacent in the periodic table. This is particularly significant given that the three known unconventional superconductors with relatively high $T_c$ involve iron, nickel, and copper. This observation suggests the possibility of discovering new unconventional superconductors based on manganese and cobalt. To explore this further, we have identified two potential manganese- and cobalt-based superconductors, with their electronic band structures illustrated in the top and middle panels of Fig. \ref{fig:a4}.

In addition to these transition metals, lithium (Li) and magnesium (Mg) also show strong potential for becoming superconductors. We speculate that these elements are more likely to be conventional superconductors when combined with non-metallic elements, similar to the well-known MgB$_2$. A promising candidate superconductor involving lithium has been identified, and its electronic band structure is depicted in the bottom panel of Fig. \ref{fig:a4}.

\nocite{*}
\bibliography{cite}
\end{document}